\input harvmac
\noblackbox


\font\cmss=cmss10
\font\cmsss=cmss10 at 7pt
\def\rlx{\relax\leavevmode}
\def\inbar{\vrule height1.5ex width.4pt depth0pt}
\def\IC{\relax\,\hbox{$\inbar\kern-.3em{\rm C}$}}
 
\def\IP{\relax{\rm I\kern-.18em P}}
\def\ZZ{\rlx\leavevmode\ifmmode\mathchoice{\hbox{\cmss Z\kern-.4em Z}}
 {\hbox{\cmss Z\kern-.4em Z}}{\lower.9pt\hbox{\cmsss Z\kern-.36em Z}}
 {\lower1.2pt\hbox{\cmsss Z\kern-.36em Z}}\else{\cmss Z\kern-.4em Z}\fi}
\def\narrowplus{\kern -.04truein + \kern -.03truein}
\def\narrowminus{- \kern -.04truein}
\def\narrowminussub{\kern -.02truein - \kern -.01truein}

\def\ep{\epsilon}

\def\rarr{\rightarrow}

\def\sqr#1#2{{\vcenter{\vbox{\hrule height.#2pt
            \hbox{\vrule width.#2pt height#1pt \kern#1pt
                  \vrule width.#2pt}\hrule height.#2pt}}}}

 
\lref\bfss{T.~Banks, W.~Fischler, S.~Shenker and L.~Susskind,
hep-th/9610043.}
 
\lref\wt{W.~Taylor IV, hep-th/9611042.}

\lref\bs{T. Banks and N. Seiberg, hep-th/9702187. }

\lref\dvv{R. Dijkgraaf, E. Verlinde, H. Verlinde, hep-th/9703030.}

\lref\tdsd{L.~Susskind, hep-th/9611164.}
 
\lref\ganor{O.~Ganor, S.~Ramgoolam and W.~Taylor, hep-th/9611202.}

\lref\vafa{C. Vafa, Nucl.Phys. {\bf B463} (1996) 415, 435.}

\lref\sena{A. Sen, Phys.Rev. {\bf D54} (1996) 2964.}

\lref\senb{A. Sen, Phys.Rev. {\bf D53} (1996) 2874.}

\lref\thft{G. 't Hooft, Comm. Math. Phys. {\bf 81}, (1981) 267.}

\lref\zram{Z. Guralnik, S. Ramgoolam, hep-th/9702099.}

\lref\vafwit{C. Vafa, E. Witten, Nucl. Phys. {\bf B431} (1994) 3.}

\lref\witbd{E. Witten, Nucl.Phys. {\bf B460} (1996) 335.}

\lref\malsuss{J. Maldacena, L. Susskind, Nucl. Phys. {\bf B475} (1996) 679.}

\lref\strovaf{A. Strominger, C. Vafa, Phys. Lett. {\bf B379} (1996) 99.}

\lref\calmald{C. Callan, J. Maldacena, Nucl. Phys. {\bf B472} (1996) 591.}

\lref\fissus{W. Fischler, E. Halyo, A. Rajaraman, L. Susskind, hep-th/9703102.}

\lref\motl{L. Motl, hep-th/9701025.}

\lref\roz{M. Rozali, hep-th/9702136.}

\lref\dvv{R. Dijkgraaf, E. Verlinde, H. Verlinde, hep-th/9703030.}

\lref\akia{A. Hashimoto, hep-th/9610250.}

\lref\akib{A. Hashimoto, W. Taylor, hep-th/9703217.}

\lref\spenta{F. Hassan, S. Wadia, hep-th/9703163.}

\lref\dasmat{S. Das, S. Mathur,  Phys.Lett. {\bf B375} (1996) 103.}

\lref\rPKT{P. K. Townsend, Phys. Lett. {\bf B350} (1995) 184.}

\lref\rWSD{E. Witten, Nucl. Phys. {\bf B443} (1995) 85.}

\lref\savsus{S. Sethi, L.Susskind,  hep-th/9702101.}

\lref\rtown{P. K. Townsend, Phys. Lett. {\bf B373} (1996) 68.}

\lref\dvva{R. Dijkgraaf, E. Verlinde, H. Verlinde, hep-th/9608096.}

\lref\bs{T. Banks and N. Seiberg, hep-th/9702187. }

\lref\dvvb{ R. Dijkgraaf, E. Verlinde, H. Verlinde, hep-th/9704018}
 
\Title{\vbox{\hbox{hep--th/9704030}\hbox{PUPT-1694}}}
{\vbox{\centerline{BPS states in Matrix Strings}}}

\smallskip
\centerline{Rajesh Gopakumar\footnote{$^\dagger$}
{gopakumr@phoenix.princeton.edu} }
\medskip\centerline{\it Department of Physics}
\centerline{\it Princeton University}\centerline{\it Princeton, NJ
08544, USA}

\vskip .3in
Matrix string theory (or more generally U-Duality) requires Super Yang-Mills 
theory to reflect a stringy degeneracy of BPS short multiplets. These are 
found as supersymmetric states in the Yang-Mills carrying (fractionated)
momentum, or in some cases, instanton number. Their energies also 
agree with those expected from M(atrix) theory. A nice parallel also emerges
in the relevant cases, between momentum and instanton number, (both integral as
well as fractional) providing evidence for a recent conjecture relating the two.  

\vskip 0.1in
\Date{2/97}

\newsec{Introduction}

In it's most recent phase of development, string theory seems to be emerging
from the confining cocoon of Riemann surfaces. While it is not clear
what it will finally metamorphose into, some features may be dimly 
discerned. One of them is the derivative nature of spacetime, perhaps 
embedded in a non-commutative space. Another is a milder exponential growth 
of the number of states as compared to the free string degeneracy. This latter 
feature, reflected in the counting of black hole microstates, is due to the 
fact that it is the BPS states that survive quantum corrections. The recent 
provocative proposal of \bfss \ tries to minimally incorporates these ideas.

But it was clear even earlier, with the identification of D-branes with 
various RR solitons, that the properties and degeneracies of BPS states must
be reflected in Super Yang-Mills theory. Through this process we have been
uncovering rather remarkable properties of Yang-Mills theory.   
In this paper we'll see a little of that in the fractionation of momentum
in field theory and its relation to instanton number.

Let us outline the problem: The conjecture made in \bfss \ essentially 
implies that the relevant degrees of freedom for M-theory on 
a transverse $T^d$, 
in the infinite momentum frame, are those of a large N Yang-Mills theory
on $T^d\times R$ with the equivalent of four dimensional $N=4$ 
supersymmetry  \wt. We shall have occasion to deal with various tori, 
but let us,
for now, consider M-theory on $T^2$. The relation with type II string theory 
on a circle implies that the $(2+1)$ dimensional Yang-Mills theory is to 
possess a whole stringy tower of states. Given the relation between 
couplings, the perturbative string spectrum is expected to 
show up only in the strong 
coupling limit of the field theory. Some arguments have recently been 
been advanced \bs \dvv \ that the non-trivial conformal field theory in this 
limit might indeed have this property. However we certainly 
expect the BPS states of the string theory to show up in the field 
theory for any value of the coupling, in particular, a semi-classical
analysis is accurate given the amount of supersymmetry. 
These states could either be ``ultra-
short'' or ``short'' depending on whether they break $1/2$ or $3/4$ of the 
supersymmetry. Given the important role that
this subset of the states play, it seems worthwhile to isolate them 
in the field theory.

The ultra-short multiplets are non-degenerate ( i.e. only 
1 multiplet with $16^2$ degrees of freedom): in the perturbative 
string spectrum they are states with both left and right moving oscillators
unexcited. As perturbative states they have purely winding or momentum modes
with respective masses, (we'll use string units $\alpha^{\prime}=1$),
\eqn\ultshrt{M^2 = (nR)^2 ; \hskip 0.5in  M^2 = ({m\over R})^2.}
Here $m$ and $n$ are the momentum and winding modes on the circle of radius
$R$. On a more general torus, these multiplets could have both winding
and momentum but in independent directions, thus for example, having a mass
\eqn\ultshort{M^2 = (n_1R_1)^2 + ({m_2\over R_2})^2.}
U-duality relates this, for instance, to an $(n_1,m_2)$ string and 
the existence of a unique 
bound state of the latter \witbd \ was a check of U-duality \sena.
These states correspond to vacua of the 
$SU(N)$ sector of the $U(N)$
Yang-Mills theory. Ultra-short multiplets were examined in the M(atrix) theory
context in \ganor, 
where their energy was found to
agree with expectations. Note that, being vacua of $SU(N)$,
the energy is purely a $U(1)$ contribution. 

The short multiplets on the other hand have an enormous degeneracy. For 
instance, they 
appear in the  perturbative spectrum on a circle with mass
\eqn\shortmass{M^2 = ({m\over R})^2 + (nR)^2 +2N_L = ({|m| \over R} + |n|R)^2.}  
Here $N_L$, the left oscillator level, is given by the level matching
condition to be $N_L = |mn|$ and $N_R=0$. The number of states is therefore given
by the number of partitions 
$d(mn)$ of level $mn$ among the physical 8 bosonic and 8 
fermionic oscillators (after taking into account the degeneracy ($16^2$) of the
ground state).
\eqn\degen{\sum d(k)q^k=16^2\prod_{m=1}({1+q^m \over 1-q^m})^8.}
The modest problem that we will tackle here, is to find these
states \foot{Note that an important difference from the previous case,
(due to the breaking of further supersymmetry)
is that these are supersymmetric BPS states in the 
$SU(N)$ Yang-Mills and not vacua. Consequently, the energy will also receive a
non-abelian contribution, but still classically computable.} in the Yang-Mills
and check that the energies and 
degeneracies match. This, it will be seen, requires the field theory to 
fractionate its momentum. (A phenomenon familiar in the black hole state
counting context \dasmat \malsuss \akia \spenta. 
See also the related recent discussion about long matrix
strings \dvva \motl \bs \dvv \ etc.) The field theory accomplishes this in an 
interesting fashion, very analogous to the fractionation of instanton
number, to which also we'll relate this phenomenon.

Actually, the fact that the degeneracies should agree is 
really a consequence of U-duality. And this exercise
may also be viewed as a check of the latter. 
We'll see, in what follows, the relation to some earlier and different checks 
of U-duality \senb\vafa \ .The energies, however, computed in the field 
theory for any finite $N$, will agree with those in M-theory only in the 
infinite momentum frame with longitudinal momentum proportional to $N$
as in \bfss. 

The next section discusses the rather special case of M(atrix) theory on 
$T^4$. Here the short multiplets
will appear in the Yang-Mills as supersymmetric states possessing fractional 
($SU(N)/Z_N$) instanton number. It will, in this sense, be a generalisation 
of the system studied by \senb \vafa and a useful preparatory case.
In Section 3 we will study the more
generic torus where the field theory fractionates it's momentum.
The short multiplets of string theory are directly visible in this case.
Section 4 returns to $T^4$ and discusses the parallel between instanton
number and momentum. We end with some discussion and conclusions. An
appendix exhibits some relevant field configurations with fractional
$SU(N)$ momentum or instanton number.   

\newsec{Magnetic Fluxes and Fractional Instanton Number}

Let's start with M(atrix) theory on $T^4$ -- it will soon be clear why we 
choose to do so. The N 0-branes, in this case,
lead a T-dualised existence as 4-branes \bfss\wt.
Thus we will examine (4+1) dimensional $U(N)$ Super-Yang-Mills 
theory on $T^4\times R$ for a stringy degeneracy of BPS states. 
The sides of the torus will be assumed to be of
length $a_i, i=1,2,3,4$. (We will use roman letters for spatial directions only.)
It was 't Hooft's observation \thft \ 
that one could have
different topological sectors corresponding to the gauge fields being
periodic upto a gauge transformation. These ``twists'' are labelled by, 
in our case, six integers $n_{ij}=-n_{ji}$ defined modulo $N$ and may
be thought of as discrete magnetic fluxes $F_{ij}$. Since
$U(N)= (SU(N)\times U(1))/Z_N$, the twist $e^{2\pi in_{ij}/N}$ in the $SU(N)$
is accompanied by one of $e^{-2\pi in_{ij}/N}$ in the $U(1)$ \zram. 

We'll consider the $SU(N)$ and $U(1)$ sectors individually, starting with
the former. We restrict ourselves to backgrounds
with static gauge fields (no scalars/fermions). Therefore we have for the 
$SU(N)$ Hamiltonian  
a bound familiar from Euclidean four dimensions:
\eqn\euclbd{H^{SU(N)}={1 \over 4g^2_{YM5}}Tr\int d^4x F_{ij}^{SU(N)}F_{ij}^{SU(N)}
\geq |{1 \over 4g^2_{YM5}}Tr\int d^4x F_{ij}^{SU(N)}\tilde{F}_{ij}^{SU(N)}|.}
It was shown in \thft \ that, in the presence of twisted boundary
conditions, that the $SU(N)$ instanton (more properly soliton)
number can be fractional, in fact,
\eqn\instno{{1\over 16\pi^2}Tr\int d^4x F_{ij}^{SU(N)}\tilde{F}_{ij}^{SU(N)}
=\nu - {\kappa \over N}.}
where $\kappa = {1 \over 4}n_{ij}\tilde{n}_{ij}= n_{12}n_{34}+n_{13}n_{42}
+n_{14}n_{23}$.
The bound on the energy is saturated for (anti) self dual fields 
$F_{ij}^{SU(N)}=
\pm \tilde{F}_{ij}^{SU(N)}$. (See appendix for some explicit configurations.)
In the supersymmetric theory at hand, it
is by now a familiar fact that, these are 
the BPS configurations since the supersymmetry variation of the $SU(N)$ gaugino
is given by 
\eqn\susyvar{\delta\chi^{SU(N)} = \Gamma^{ij}F_{ij}^{SU(N)}\ep.}
A (anti) self-dual field implies that for half of the $\ep$'s (those 
of the appropriate chirality), the variation vanishes. 

Let us remain in a sector where the integer component $\nu$ is zero.
Moreover, for simplicity, we'll also take only $n_{12}, n_{34}$ non-zero.
(This is also the (422) system studied in \zram.) 
We then have a BPS state with energy given by   
\eqn\hsun{H^{SU(N)}={4\pi^2 \over g^2_{YM5}N}|n_{12}n_{34}|}

There is a $U(1)$ contribution to the energy too. This is a result of the
identification of the twists of the $U(1)$ and the $SU(N)$. We have 
constant $U(1)$ magnetic fluxes 
\eqn\mflux{F_{ij}^{U(1)}a_ia_j = {2\pi n_{ij}\over N}}
with no sum over indices. (These $U(1)$ fluxes give an opposite fractional 
contribution to the total $U(N)$ instanton number so that the net number is zero
\zram.) 
This does not break any further supersymmetries
since 
\eqn\susyvarui{\delta\chi^{U(1)} = \Gamma^{ij}F_{ij}^{U(1)}\ep +\ep^{\prime},}
and one can choose $\ep^{\prime}$ to cancel the $\ep$ variation. 
The $U(1)$ part of the energy is therefore
\eqn\hui{H^{U(1)}= {1\over 2g^2_{YM5}}\int d^4x ((F_{12}^{U(1)})^2 
+(F_{34}^{U(1)})^2)
={2\pi^2\over g^2_{YM5}N}
({a_3a_4\over a_1a_2}n_{12}^2 +{a_1a_2\over a_3a_4}n_{34}^2).}
The total energy is now
\eqn\hun{H^{U(N)}=H^{U(1)}+H^{SU(N)}={2\pi^2a_1a_2a_3a_4\over g^2_{YM5}N}
({|n_{12}|\over a_1a_2} +{|n_{34}|\over a_3a_4})^2.}

What about the degeneracy of a state with this energy? The moduli space of
instantons in $SU(N)/Z_N$ gauge theory with instanton number $\nu$ is believed
to be $(T^4)^{N\nu}/S(N\nu)$ \vafa. This is presumably true for both 
fractional 
and integral instanton number $\nu$ since one can separate a configuration with
integral instanton number into clusters with fractional charge.
Quantising the collective coordinates
and looking at the ground states of the corresponding 
supersymmetric quantum mechanics, one finds the degeneracy $d(N\nu)$ 
to be given by the 
dimension of the cohomology of the above orbifold. It's generating
function has been calculated in \vafwit \ to be precisely \degen . (The factor
of $16^2$ comes from the $U(1)$ sector -- it was present in the case of
short multiplets as well.)
For a fractional instanton number $\nu= n_{12}n_{34}/N$, the degeneracy is
$d(n_{12}n_{34})$. 

Now for some comparisons. The configuration we have considered is in the 
language of 4-branes, a collection of $N$ 4-branes together with 
$n_{12}, n_{34}$ 2-branes 
wrapped on directions $(34)$ and $(12)$ respectively. 
\foot{The system of 2-branes at angles has also recently been studied from the
point of view of fractional instantons \akib }
This follows from
the fact that fluxes $F_{12}$ and $F_{34}$ act as sources of 2-brane 
charge in the 4-brane world volume. This is a short multiplet configuration
which is U-dual to the perturbative state with
momentum $n_{12}$ and winding $n_{34}$, both in, say, direction 2.
(This may be seen on following this configuration through the 
dualities $T_1ST_{1234}ST_1$. $T_i$ denotes T-duality in direction $i$.
The 4-branes go over into winding $N$ in the 1 direction.)
It is satisfying that we saw precisely the degeneracy $d(n_{12}n_{34})$
in the gauge theory that we expect for this short state in string theory. 

This is closely related to the results of \senb\vafa. They consider
in the system of $N$ 4-branes, $m$ 0-branes instead of the system of 2-branes
that we had above. In the gauge theory, this corresponds to the sector 
with integral instanton number $m$. This is U-dual via the same duality chain
in the previous paragraph to winding $N$ and momentum $m$ in direction 1. 
The degeneracy is thus $d(Nm)$ which is also the degeneracy of states
with integral instanton number $m$ in the $SU(N)$ theory. We can now easily see
that it is possible to generalise to the case of arbitrary instanton
number by combining 0-branes and 2-branes. The degeneracies are precisely those
expected of a string state with winding and momentum simultaneously in
directions 1 and 2.

In the M(atrix) theory picture, our gauge configuration is (the T-dual of)
the configuration with $n_{12}, n_{34}$ 2 branes wrapped on directions 
$(12)$ and $(34)$ respectively, with $N$ 0-branes as always. This follows, 
again, from 
the identification of magnetic fluxes, $F_{12}$ and $F_{34}$ with 2-branes
\ganor. As we saw, having only fractional instanton charge means, 
in the language of 4-branes, that there are no 0-branes. Or equivalently,
no longitudinal 5-branes in M(atrix) theory. 

We'll compare the energy \hun \ in gauge theory with that expected of
this configuration of branes in M-theory. 
The mass of this
bunch of wrapped 2-branes in M-theory is (written in terms of type II string 
parameters)  
\eqn\twobr{M={(|n_{12}|l_1l_2 +|n_{34}|l_3l_4)\over g_s},}
where $l_i$ are the radii of the $S^1$'s. To relate to the gauge
theory variables recall that (by T-duality \wt \bfss) $l_i={2\pi \over a_i}$.
We also need the identification of the (open) string coupling 
with the gauge coupling
\eqn\coupl{{1\over g^2_{YM5}}= (2\pi)^2{l_1l_2l_3l_4\over g_s};
\hskip 0.5in g_s=R_{11}}
and the fact that the energy in the infinite momentum frame is given by
\eqn\infm{E={M^2\over 2p_{11}}; \hskip 0.5in p_{11}={N\over R_{11}}.}
Putting it all together 
gives precisely the energy of the state in gauge theory \hun. 

A few words should be said about ultrashort configurations. If we had 
instead turned
on fluxes, say, $n_{12}, n_{24}$, then we would not have had any fractional
instanton number. We would have had zero energy from the $SU(N)$ part --
in other words, a unique vacuum state. All the energy would have come from
\hui (with index 3 replaced by 2). It is easy to check that the set of 2-branes 
this corresponds to, (wrapped on (34) and (13) directions) is U-dual to winding
and momentum along different directions, an ultrashort state. The energy of this
state in M-theory in the infinite momentum frame is also in agreement with 
\hui ,(compare with \ultshort ).

\newsec{Electric, Magnetic fluxes and Momentum}

Thus far we have been on a special torus which admitted instantons and had
switched on only the 't Hooft magnetic fluxes. We are now ready to 
consider a more general case. For definiteness, we'll mostly deal with 
M(atrix) theory on $T^2$, though the generalisations will be evident. 
The relevant gauge theory is $2+1$ dimensional with gauge group $U(N)$.
We will also be turning on discrete electric fluxes. Physically, this is 
equivalent to the presence of a Wilson line, which creates the electric flux,
winding along one of the compact directions. The distinct topological sectors
are labelled by integers $q_i$ defined again modulo $N$ corresponding to
Electric fluxes $F_{0i}$.
  
As before, we first consider the $SU(N)$ sector. We do not have an instanton
bound, but rather the simpler (but less utilised)
\eqn\minkbd{H^{SU(N)} \geq |P_i| \equiv |\int d^2x T_{0i}| =
{1\over g^2_{YM3}}Tr|\int d^2x F_{0i}^{SU(N)}F_{ij}^{SU(N)}|.}
The left hand side is simply the norm of the Non-abelian Poynting vector which
measures the momentum in the field. Consider a topological sector with 
$n_{12},q_2 \neq 0$ and oscillator momentum $m_1$ in direction 1. Then 
\eqn\topmom{ H^{SU(N)} \geq {1\over g^2_{YM3}}Tr|\int d^2x 
F_{0i}^{SU(N)}F_{ij}^{SU(N)}| = {4\pi^2\over a_1}|m_1 - {n_{12}q_2\over N}|.}
It is no surprise to see the contribution ${4\pi^2m_1\over a_1}$ to the momentum, 
(Note that $a_i$ are the circumferences, not the radii) but
the fractional term is not usual. However, it is there since one can write
down explicit field configurations analogous to those with fractional 
instanton number (see appendix). And while one does not think of momentum
as a topological quantum number, on a compact domain the quantisation
makes it impossible for this number to change continuously on varying
the field configuration. We thus see clearly in the field theory context how
fractional momentum emerges. It is the momentum associated with the discrete
fluxes that a $SU(N)/Z_N$ gauge theory admits. 

When is the bound on the energy satisfied? In the case we have been considering
it occurs when
\eqn\bound{F_{02}^{SU(N)}=\pm F_{12}^{SU(N)} \equiv \hat{F}.}
This may look unfamiliar but it is just the non-abelian generalisation of
an electromagnetic plane wave travelling in direction 1. The electric and magnetic
fields are orthogonal (in directions 2 and 3, say) and have equal norm.
In our supersymmetric theory, it is nice fact that these are also BPS states.
We see that \susyvar \ reduces to 
\eqn\elsusy{\delta\chi^{SU(N)}\propto \Gamma^2(\Gamma^0\pm\Gamma^1)\ep\hat{F} .}
The ``Dirac equation'' structure immediately tells us that for 
half of the $\ep$'s, the variation vanishes. 

In analogy with the previous section, let us remain in the sector with 
$m_1$ zero, i.e. with only fractional $SU(N)$ momentum ${n_{12}q_2\over N}$. 
(We must remind the reader that just as with instanton number, the nett $U(N)$
momentum is always integral. But the spacing between levels is in units of 
${1\over N}$.)
Then
\eqn\hsunel{H^{SU(N)}={4\pi^2 |n_{12}q_2|\over Na_1}}
The $U(1)$ contribution arises from the fact that the two sectors are correlated.
As before we have 
\eqn\minkmflux{F_{12}^{U(1)}a_1a_2 = {2\pi n_{12}\over N}.}
But now there is an Electric flux as well whose quantisation condition reads as 
\eqn\minkeflux{E_2^{U(1)}a_1={2\pi q_2\over N}; \hskip 0.5in 
E_2^{U(1)} \equiv {1\over g^2_{YM3}}F_{02}^{SU(N)}}
Thus 
\eqn\huiel{H^{U(1)}= {g^2_{YM3}\over 2}\int d^2x (E_2^{U(1)})^2
+{1\over 2g^2_{YM3}}
\int d^2x (F_{12}^{U(1)})^2= {2\pi^2g^2_{YM3}\over N}q_2^2({a_2\over a_1})+
{2\pi^2n_{12}^2\over g^2_{YM3}Na_1a_2}.}
The total energy now reads as 
\eqn\hunel{H^{U(N)}=H^{U(1)}+H^{SU(N)}={2\pi^2g^2_{YM3}\over a_1a_2N}
(q_2a_2 +{n_{12}\over g^2_{YM3}})^2}
Now for the degeneracy of these momentum states. Here, we do not yet have a 
complete argument. But it is plausible, given that momentum is quantised in
units of ${1\over N}$, that the degeneracy from the $SU(N)$ sector 
will be given by the number 
of ways $d(n_{12}q_2)$ of partitioning $n_{12}q_2$ amongst 8 bosonic and 
fermionic modes. 
This is similar to the counting in \malsuss.   

We are now ready to make some comparisons. We have been essentially 
considering  
the world volume theory of N 2-branes together with $n_{12}$ 0-branes and 
fundamental strings with winding $q_2$. This follows from the identification
\witbd \ of electric fluxes with winding of fundamental strings.
Again, this is a short multiplet configuration
which is U-dual to a perturbative string state with momentum $n_{12}$ and winding
$q_2$, both in direction 2. (As follows from the duality chain $T_2ST_2$. The 
2-branes go over into strings with winding  $N$ in direction 1.) 
The degeneracy of this latter state
is $d(n_{12}q_2)$, in accord with what we had in the Yang-Mills. The case with 
only integer momenta $m_1$ in \topmom , is U-dual, via the same chain, to 
$m_1$ units of momentum together with winding $N$ in direction 1, thus another short
multiplet configuration. Now the degeneracy is $d(Nm_1)$, once again something that
might be expected in the Yang-Mills given the fractional units of momentum.
Again, the case of arbitrary momentum in \topmom poses no problems, being 
the combination of the above two cases and has the right degeneracy. All this
is a test of U-duality for M-theory on tori, which is independent from the one
performed in \senb \vafa , as we'll elaborate a bit in the next section. 

On a general torus $T^d$, the considerations are very analogous. The $U(N)$
Yang-Mills is the world-volume theory of N $d$ branes. One may have $n_{12}$
$(d-2)$ branes wrapped on directions $(3,\ldots ,d)$ as the source of magnetic flux
and fundamental strings with winding $q_2$ on direction 2 as that of electric flux.
Integer momenta $m_1$ may also be present. This would be dual (via 
$T_{2\ldots d}ST_{2\ldots d}$ for even $d$ and $T_{1\ldots d}ST_{1\ldots d}$ for odd
$d$) to the short string multiplets of the previous paragraph. The counting works 
exactly as before. Only the mass formula has some modifications for the dimensionality.
It is clear, for instance,  that we will need to consider the (5+1) dimensional 
Field theory with
configurations of both instantons and momenta to make contact with the states
responsible for Black Hole entropy \strovaf \calmald.

Coming to the M(atrix) theory on $T^2$ interpretation, our original 
gauge configuration is (the T-dual of) a collection of N 0-branes, $n_{12}$ transverse
2-branes wrapped on the $T^2$ and $q_2$ units of Kaluza-Klein momentum in direction 2.
The case with momentum $m_1$ corresponds to a longitudinal 2-brane wrapped on  
direction 1. We can compare the energy of this set-up with \hunel. The mass is
given by (taking $m_1 =0$)
\eqn\minkmass{M= {|n_{12}|l_1l_2\over g_s} +{|q_2|\over l_2}.}
As before, $l_i={2\pi\over a_i}$. The gauge coupling is now given by
\eqn\gaugecoup{{1\over g^2_{YM3}}= {2\pi l_1l_2\over g_s}.}
Together with \infm we reproduce exactly \hunel.

There is another way of putting this so that we can directly see how gauge 
theory reproduces the perturbative string spectrum \shortmass. Running the argument 
backwards, we could say that the Yang-Mills calculation \hunel \
 predicts, using
the usual identifications \gaugecoup and \infm , that the mass of a state in M-theory 
on $T^2$ with $n_{12}$ 2-branes and KK momentum $q_2$ is
\minkmass. Now M-theory on $T^2$ is equivalent to
string theory on $S^1$, taking say, direction 1 to be the ``eleventh'' one. The 
$n_{12}$ 2-branes are, from this angle, fundamental strings with winding $n_{12}$ on
direction 2. The tension of this string is given by 
${1\over 2\pi\alpha^{\prime}}={l_1\over 2\pi g_s}$ -- derived from the 
2-brane wrapped on direction 1, (using the 
2-brane tension derived in M(atrix) theory \bfss). 
The KK momenta $q_2$ in direction 2 remain what they are. We immediately see,
with this identification of the tension, and $l_2\rarr R, n_{12}\rarr n, q_2\rarr m$ 
that \minkmass coincides with \shortmass.

\newsec{Instantons and Momenta} 

The development of  the last two sections has been rather analogous. Instantons
and momenta are both fractionated in units of ${1\over N}$, compare equations
\instno \ and \topmom . However the instantons appeared only in the case of $T^4$,
while momenta are present in any dimension! So, let us focus, once more, on the case of
(4+1) dimensional $SU(N)$ Super Yang-Mills where they both occur, and 
examine the correspondence a bit better. 

Let's take integer $SU(N)$ quantum numbers 
(though the conclusions would be the same for fractional charges) 
for momenta/instanton number.
Thus the two physical situations are : 1. $N$ 4-branes with $m$ 0-branes, 
2. $N$ 4-branes with momentum $m$ in direction 1. There is a third configuration
that they are both U-dual to, namely, the string state with winding $N$ and momentum
$m$ in direction 1. These represent the three ``conjugacy classes'' for short
multiplets under the U-duality group for String theory on $T^4$. 
\foot{Any two configurations equivalent upto T-duality (i.e. perturbatively
equivalent)
belong to the same ``conjugacy class''. There are two such classes of short multiplets
for string theory on $T^d$, $d<4$ and three for $d=4$. This is an equivalent reason
as to why $T^4$ is special.}
It is the first two that we have considered as BPS states in the 
Yang-Mills. We have seen that they have exactly the same degeneracy.
In fact, the form of \instno \ and \topmom , suggest
an interpretation of $g^2_{YM5}$ as the size of a fifth circle, on the same footing
as the $a_i$. Instantons (actually solitons) are then momenta in this direction.
The fractionation of momenta might be used to motivate that of instanton number or
vice versa. In fact, the identification of (integral) instanton number with momenta 
and the emergence of a new dimension is precisely the suggestion made recently 
in \roz (see also \fissus) .
We can now give some more evidence for this conjecture. We have seen that there is a 
whole tower of BPS states which transform into each other under the  postulated 
$SL(5,Z)$ symmetry. Moreover, their energies also reflect this invariance. 
If we redo the calculations \minkeflux \huiel \hunel in the (4+1) dimensional
case, we obtain for the energy of a state with $n_{12},q_2 \neq 0$
\eqn\hunelfour{H^{U(N)}={2\pi^2 a_2a_3a_4g^2_{YM5}\over a_1N}({|q_2|\over a_3a_4}
+{|n_{12}|\over g^2_{YM5}a_2})^2}
Comparing with \hun \ , with the identification $g^2_{YM5}\equiv a_5$, we 
immediately see the symmetry among the ``five'' directions. 
We also see that the Poynting vector and Instanton density are supposed to transform
under the  postulated $SL(5,Z)$ symmetry of the theory. It might perhaps be possible
to combine them into a symmetric form.

\newsec{Discussion and Conclusions}

The objective of this work was to find, purely within Yang-Mills theory a description
of the BPS states of string theory. 
It is rather surprising that Yang-Mills theory seems to possess non-trivial
characteristics which are demanded by string theory \savsus \bs . Here we have seen 
a hagedorn number of BPS states, carrying both 
momentum {\it and} winding quantum numbers and having the energies expected
of the string spectrum.
This was achieved by the field theory fractionating its momentum in a manner 
appropriate to long strings. This is done in a way so similar to that of instanton
number that in the right circumstances, the interchange of instantons with momentum
becomes a symmetry -- one demanded once again by string theory. Of course, since 
this could be phrased in the context of M(atrix) theory,
we have equivalently made some new checks of U-duality and M(atrix) theory. Our study
have been for arbitrary $N$, but the energies 
coincide with that of M-theory/string theory
only in the infinite momentum frame with the identification of the longitudinal momentum
$P_{11}=N/R_{11}$. This is a general feature of toroidal compactifications of
M(atrix) theory (see remarks in \fissus). Finally, the presence of a U-dual spectrum 
of states in the Yang-Mills is also a signature of Lorentz invariance.

\bigskip
{\bf Acknowledgements:}
I must specially thank Savdeep Sethi for long, useful conversations on the topic.
It is also a pleasant task to acknowledge discussions with O. Ganor, D. Gross,
A. Hashimoto and S. Ramgoolam. I must also thank the I.T.P. Santa Barbara, where
a part of this work was done, for hospitality. This research was supported
in part by the National Science Foundation under Grant NO. PHY94-07194.   

\bigskip
{\bf Note Added:}
While this manuscript was just being completed, we received the preprint \dvvb
in which BPS states in the matrix theory were used to study black hole issues in 
5 dimensions.

\bigskip
{\bf Appendix}

Following \thft and \zram we'll display some field configurations 
with fractional $SU(N)$ instanton number and momentum. We'll see that they are
very similar. More details can be had from these references. 

For a (4+1) dimensional $U(N)$ gauge theory with twists $n_{12},n_{34}\neq 0$,
the relevant fields are, for a self dual $SU(N)$ configuration, 
\eqn\fsun{F_{12}^{SU(N)}={F\over N}\omega = F_{34}^{SU(N)}}
where $\omega=diag(\overbrace{k,\ldots ,k}^{l\, times},
\overbrace{-l,\ldots ,-l}^{k\, times})$ is the generator of the $U(1)^{\prime}$ in 
$SU(l)\otimes SU(k)\otimes U(1)^{\prime}\subset SU(N)$ with $k+l=N$. $F$ is a constant 
determined by 
\eqn\Fcond{Fa_1a_2= {2\pi n_{12}\over l}; \hskip 0.5in Fa_3a_4=-{2\pi n_{34}\over k},}
which also determines $k,l$ in terms of the other parameters. 
The correlation of the twists between the $SU(N)$ and $U(1)$ parts imply that the 
$U(1)$ field strengths are determined to be 
\eqn\fui{F_{12}^{U(1)}={lF\over N}{\bf 1}_{N\times N}; \hskip 0.5in F_{34}^{U(1)}=
-{kF\over N}{\bf 1}_{N\times N}.}
Note that the $U(1)$ field strengths are not self-dual.
We see that the $SU(N)$ and $U(1)$ instanton numbers are opposite in sign and equal
in magnitude to ${n_{12}n_{34}\over N}$. 

The only modification in the case with electric fluxes is in some quantisations.
If, as in Section 3., $n_{12},q_2 \neq 0$, then again
\eqn\felsun{F_{12}^{SU(N)}={F^{\prime}\over N}\omega = F_{02}^{SU(N)}.}
with 
\eqn\fprime{F^{\prime}a_1a_2 = {2\pi n_{12}\over l} \hskip 0.5in 
{1\over g^2_{YM3}}F^{\prime}a_1 =-{2\pi q_2\over k}.}
The $U(1)$ fields are 
\eqn\felui{F_{12}^{U(1)}={lF^{\prime}\over N}{\bf 1}_{N\times N}; \hskip 0.5in 
F_{34}^{U(1)}=-{kF^{\prime}\over N}{\bf 1}_{N\times N}.}

\listrefs  
\end